\documentclass[english]{cccconf}

\usepackage{amsmath}
\newtheorem{thm}{Theorem}

\newtheorem{lem}{Lemma}
\newtheorem{remark}{Remark}

\begin{document}

\title{Coordination for a Group of Autonomous Mobile Agents with Multiple Leaders}
\author{Jiangping Hu, Yiguang Hong}

\affiliation{Institute of  Systems Science,
        Chinese Academy of Sciences, Beijing 100080
        \email{jphu@amss.ac.cn,yghong@iss.ac.cn}}
\maketitle

\begin{abstract}
In this paper, we consider the coordination control of a group of
autonomous mobile agents with multiple leaders. Different
interconnection topologies are investigated. At first, a necessary
and sufficient condition is proved in the case of fixed
interconnection topology. Then a sufficient condition is proposed
when the interconnection topology is switched. With a simple
first-order dynamics model by using the neighborhood rule, both
results show that the group behavior of the agents will converge
to the polytope formed by the leaders.
\end{abstract}


\footnotetext{IEEE Catalog Number: 06EX1310}
 \footnotetext{This work was supported by the NNSF of China under Grants 60425307, 50595411 and 60221301.}

\section{INTRODUCTION}

Recent years have seen a large and growing literature concerned
with the coordination of a group of autonomous agents, partly due
to a broad application of multi-agent systems including
flocking/swarming (e.g., \cite{rey87, vic95}), formation control
(e.g.,\cite{fax,lin05}), and sensor networks (e.g., \cite{cort}).
Leader-following problem is one of the important coordination
problems in the studies of multiple mobile agents.  Even in some
leaderless cases, concepts like ``virtual leader" are proposed to
study cooperative behaviors \cite{shi, saber06}.

Usually, there is only one leader in the leader-following
formulation, though sometimes, the leader may be active with
unmeasurable states \cite{hong}.  However, in some practical
situations of formation or foraging, the formulation of multiple
leaders may be needed. In \cite{lin05}, Lin {\it et al.} discussed
an interesting model for a group of agents with straight-line
formation containing two "edge leaders", where all the agents
converge to a uniform distribution on the line segment specified
by the two edge leaders. In \cite{iain}, a simple model was given
to simulate foraging and demonstrate that, the larger the group
is, the smaller the proportion of ``leaders" is needed to guide
the group.

Inspired by \cite{lin05,iain}, in this paper we consider the
coordination behavior of mobile agents with multiple leaders. By
using a neighborhood rule, we show that a group of agents will
converge to the polytope formed by the leaders (that is, the
leaders forms the vertex set of the polytope) in two different
cases of interconnection topologies associated with the agents and
the leaders, and demonstrate that the collective behavior changes
as the connectivity of the interconnection between agents and some
leader increases through some simulations.

This paper is organized as follows.  Section 2 presents the
problem formulation for multiple leaders. Then, both fixed
interconnection topology and switched topology are considered, and
the corresponding coordination behaviors are analyzed in Sections
3 and Section 4. In Section 5, two numerical examples are shown.
Finally, concluding remarks are given in Section 6.

\section{PROBLEM FORMULATION}

Before formulating our problem, we first introduce some basic
concepts and notations in graph theory that will be used
\cite{god}.

Let $\mathcal{G}=(\mathcal{V},\mathcal{E},A)$ be a weighted
undirected graph of order $n$ with a set of nodes
$\mathcal{V}=\{1,2,...,n\}$,  set of edges $\mathcal{E}\subseteq
\mathcal{V} \times \mathcal{V}$, and a weighted adjacency matrix
$A=[a_{ij}]\in \mathcal{R}^{n \times n}$ with nonnegative
elements.  The node indexes belong to a finite index set
$\mathcal{I}=\{1,2,...,n\}$. An edge is denoted by $(i,j)$, which
means node $i$ and node $j$ are adjacent. A path from $i$ to $j$
in $\mathcal{G}$ is a sequence of distinct nodes starting with $i$
and ending with $j$ such that the consecutive nodes are adjacent.
If there is a path between any two nodes of a graph $\mathcal{G}$,
then $\mathcal{G}$ is connected, otherwise disconnected. An
induced subgraph ${\mathcal X}$ of $\mathcal{G}$ that is maximal,
subject to being connected, is called a component of
$\mathcal{G}$. The element $a_{ij}$ associated with the edge of
the graph is positive, i.e. $a_{ij}>0 \Leftrightarrow (i,j) \in
\mathcal{E}$. Moreover, we assume $a_{ii}=0$ for all $i \in
\mathcal{I}$. The set of neighbors of node $i$ is denoted by
$\mathcal{N}_{i}=\{j\in \mathcal{V}:(i,j)\in \mathcal{E}\}$.

A diagonal matrix $D=diag\{ d_1,...,d_n\}\in R^{n\times n}$ is a
degree matrix of $\mathcal{G}$, whose diagonal elements
$d_i=\sum_{j \in \mathcal{N}_{i}}a_{ij}$ for $i=1,...,n$. Then the
Laplacian of the weighted graph $\mathcal{G}$ is defined as
$L=D-A\in R^{n\times n}. $ The following result is well-known in
algebraic graph theory (e.g. \cite{god}) and establishes a direct
relationship between the graph connectivity and its Laplacian.

\begin{lem}
\label{lem1} Let $\mathcal{G}$ be a graph on $n$ vertices with
Laplacian $L$. Denote the eigenvalues of $L$ by
$\lambda_{1}(L),\cdots, \lambda_{n}(L)$ satisfying
$\lambda_{1}(L)\leq\cdots\leq\lambda_{n}(L)$.  Then
$\lambda_{1}(L)=0$ and $\textbf{1}=[1,1,\cdots,1]^{T}\in R^n$ is
its eigenvector. Moreover, if ${\mathcal G}$ is connected,
$\lambda_2>0$.
\end{lem}

Next, we introduce some notations in convex analysis \cite{rock}
and linear algebra \cite{horn}.  Let $S \subset R^m$. $S$ is said
to be convex if $(1-\gamma)x+\gamma y\in S$ whenever $x\in S,y\in
S$ and $0<\gamma <1$. A vector sum $\gamma_1 x_1+\cdots+\gamma_n
x_n$ is called a convex combination of $x_1,\cdots,x_n$ if the
coefficients $\gamma_i$ are all non-negative and
$\gamma_1+\cdots+\gamma_n=1$. Here, $\gamma_i$ can be interpreted
as proportions.  The intersection of all convex sets containing
$S$ is the convex hull of $S$, denoted by $co(S)$. The convex hull
of a finite set of points $x_1,\cdots,x_n\in R^m$ is a polytope,
denoted by $co\{x_1,\cdots,x_n\}$. If $x_2-x_1,\cdots,x_n-x_1$ are
linearly independent, the set of points $x_1,\cdots,x_n$ is said to be
affinely independent. Then the polytope is called an $n-1$ dimensional
simplex and  $x_1,\cdots,x_n$ are called the vertices of the simplex.

In this paper, we consider a system consisting of $n$ agents and
$k$ leaders, and the interconnection topology among them can
easily be described by a simple graph.  The purpose of the leaders
is to guide the multi-agent behavior. Denote the positions of
these static leaders by $x_0^j \in R^m,j=1,\cdots,k$. The
interconnection topology $\mathcal{G}$ associated with $n$ agents
can be fixed or variable but may not be connected.  If we regard
the polytope formed by leaders as a virtual node and if one agent
can see a vertex (i.e. some leader) of the polytope, we say
that the agent is connected to the virtual node. By "the graph,
$\bar{\mathcal G}$, of this
system is connected", we mean that at least one agent in each
component of ${\mathcal G}$ is connected to the virtual node. Then
we define a diagonal matrix $B$ to be a leader adjacency matrix
associated with $\bar{\mathcal{G}}$ with diagonal elements
$b_i\;(i\in \mathcal{I})$ such that each $b_i$ is some positive
number if agent $i$ is connected to the virtual node and $0$
otherwise.

A continuous-time dynamics of $n$ agents is described as follows:
\begin{equation}
\label{eq0}  \dot {x}_i = u_i,
\end{equation}
where $x_i\in R^m$ can be the position of agent $i$ and $u_i\in
R^m$ its interconnection control inputs for $i=1,...,n$.  As
usual, we propose a neighbor-based feedback control as follows:
\begin{equation}
\label{con}
\begin{split}
u_i=\sum_{j\in N_i(\sigma)}&a_{ij}(x_j-x_i)+\sum_{q=1}^k b_i^q(\sigma)(x_0^q-x_i);\\
 &i=1,\cdots,n.
\end{split}
\end{equation}
where nonnegative functions $b_i^q(\sigma(t))>0$ if and only if
agent $i$ is connected to leader $q\; (q=1,\cdots,k)$ when $t\in
[t_l,t_{l+1})$ for $l=0,1,\cdots$, and switching signal $\sigma:
[0,\infty) \to \mathcal{P}=\{1,...,N\}$ ($N \in Z^{+},$ denotes
the total number of all possible digraphs) is a switching signal
that determines the interconnection topology $\bar{\mathcal{G}}$.
If $\sigma$ is a constant function, then the corresponding
interconnection topology is fixed.

Denote
$$
x=\left(\begin{array}{c}x_1\\\vdots\\x_n\end{array}\right)\in
R^{mn},
x_0=\left(\begin{array}{c}x_0^1\\\vdots\\x_0^k\end{array}\right)\in
R^{km}.
$$
Let
$B_\sigma^q=diag\{b_1^q(\sigma(t)),\cdots,b_n^q(\sigma(t))\}\in
R^{n \times n}$ be a diagonal matrix with non-negative diagonal
entry $b_i^q(\sigma(t))$ for $q=1,\cdots,k$ and
$B_\sigma=[B_\sigma^1,\cdots,B_\sigma^k]\in R^{n\times nk}$. Let
$\Xi=\{\xi=(\xi_1,\cdots,\xi_n)^T|\xi_i\in
co\{x_0^1,\cdots,x_0^k\}\}$.

Then we rewrite the closed-loop system in a compact form:
\begin{equation}
\label{eq1}
\dot {x}\in F(x),
\end{equation}
with a set-valued function
$$
F(x)=\{-(H_p \otimes I_m) x+[B_p(I_k\otimes \textbf{1}_n)]\otimes I_m x_0|p\in \mathcal{P}\}
$$
and
$$
H_p=L_p+B_p(\textbf{1}_k\otimes I_n),
$$
where $\otimes$ is the Kronecker product.

Throughout the paper, $R^+=[0,\infty),||\cdot||$ denotes the
Euclidean norm and $<\cdot,\cdot>$ denotes inner product on
$R^{mn}$. For non-empty $\Xi\subset R^{mn}$, $d_\Xi :
R^{mn}\rightarrow R^{+}$ denotes its Euclidean distance function
given by $d_\Xi(x)=\frac{1}{2}\underset{\xi\in
\Xi}{inf}||x-\xi||^2$.

The objective of this work is to lead all the agents to enter the
region formed by the leaders; namely, $x_i\;(i=1,\cdots,n)$, for
each agent $i$, will be contained in a convex hull of
$x_0^j\;(j=1,\cdots,k)$ as $t\to \infty$.  In other words,
\begin{equation}
\label{aim} \lim_{t\to \infty} d_\Xi(x(t))= 0.
\end{equation}
In the following sections, the convergence of the system
(\ref{eq1}) will be studied based on $d_\Xi(x)$, with either fixed
interconnection topology or switched interconnection topology.

\section{FIXED INTERCONNECTION TOPOLOGY}

In this section, we will focus on the convergence analysis of a
group of dynamic agents with fixed interconnection topology.  In
this case, the subscript $\sigma$ can be dropped.

Then the differential inclusion (\ref{eq1}) becomes
\begin{equation}
\label{eq2}
\dot {x}=-(H \otimes I_m) x+[B(I_k\otimes \textbf{1}_n)]\otimes I_m x_0.
\end{equation}

The next lemma shows the relationship between the positive
definiteness of a matrix $H=L+B(\textbf{1}_k\otimes I_n)$ and the
connectivity of $\bar{\mathcal{G}}$, which was proved in
\cite{hong}.

\begin{lem}
\label{lem2} If graph $\bar {\mathcal G}$ is connected,  then the
symmetric matrix $H$ associated with $\bar {\mathcal G}$ is
positive definite.
\end{lem}

Then a main result associated with the fixed interconnection
topology is given as follows:

\begin{thm}
\label{thm1} For the system (\ref{eq2}), (\ref{aim}) holds if and
only if $\bar{\mathcal{G}}$ is connected,
\end{thm}

Proof: (Sufficiency) Let $\bar x=x-[H^{-1}B(I_k\otimes
\textbf{1}_n)]\otimes I_m x_0$.   Then, from equation (\ref{eq2}),
we have
\begin{equation}
\label{eq3}\dot{\bar{x}}=-(H\otimes I_m)\bar x.
\end{equation}
For any initial value $\bar x(t_0)$ at initial time $t_0$, the solution of the system (\ref{eq3}) is
$$
\bar x(t)=e^{-(H\otimes I_m)(t-t_0)}\bar x(t_0),\; t\geq t_0.
$$
From Lemma \ref{lem2}, the eigenvalues of $H\otimes I_m$ are
positive, and therefore, $ \bar x(t) \to 0 $; namely, $x \to
x^*=[H^{-1}B(I_k\otimes \textbf{1}_n)]\otimes I_m x_0$, as $t\to
\infty$.

Then we only need to prove that each vector $x_i^*\in
R^m\;(i=1,\cdots,n)$ can be represented by a convex combination of
$x_0^p\in R^m$ for $p=1,\cdots,k$. It is equivalent to prove that
$[H^{-1}B(I_k\otimes \textbf{1}_n)]\otimes I_m$ is a row
stochastic matrix which is a non-negative matrix and the sum of
the entries in every row equals $1$.

For $H$, there exists a positive number $\alpha$ such that
$H=\alpha I-M$, where $M$ is a non-negative matrix and
$\alpha>\rho(M)$ with $\rho(M)=\max\{\sqrt{\lambda},\lambda\;\mbox
{is an eigenvalue of} \;M^TM\}$. In fact, the eigenvalue
$\lambda_i(H)=\alpha-\lambda_i(M)>0,i=1,\cdots,n$. Thus,
$$
(\alpha
I-M)^{-1}=\frac{1}{\alpha}(I+\frac{1}{\alpha}M+\frac{1}{\alpha^2}M^2+\cdots)\geq
0_{n\times n}
$$
or equivalently, $(H^{-1}\otimes I_m)$ is a non-negative matrix,
and so is $[H^{-1}B(I_k\otimes \textbf{1}_n)]\otimes I_m$.

Additionally, since
$$
(H\otimes I_m)(\textbf{1}_n\otimes
\textbf{1}_m)=(\sum_{q=1}^{k}B^q\textbf{1}_n)\otimes \textbf{1}_m,
$$
we have
\begin{equation*}
\begin{split}
&(H\otimes I_m)^{-1}(\sum_{q=1}^{k}B^q\textbf{1}_n)\otimes
\textbf{1}_m\\
 =&((H\otimes I_m)^{-1}[B(I_k\otimes \textbf{1}_n)]\otimes I_m)(\textbf{1}_k\otimes
\textbf{1}_m)\\
=&([H^{-1}B(I_k\otimes \textbf{1}_n)]\otimes I_m)(\textbf{1}_k\otimes
\textbf{1}_m)=\textbf{1}_n\otimes \textbf{1}_m.
\end{split}
\end{equation*}

Note that $[H^{-1}B(I_k\otimes \textbf{1}_n)]\otimes I_m$ is a
matrix with every row sum equal to $1$, which leads to the
conclusion.

(Necessity) Suppose that $\bar{\mathcal{G}}$ is not connected.
Without loss of generality, we also suppose that there are
$\kappa$ components in $\mathcal{G}$, where some components are
connected with the virtual node. We can renumber the nodes of
$\mathcal{V}$ such that Laplacian matrix associated with
$\mathcal{G}$ takes the following form
\begin{equation*}
L=\left(\begin{array}{cc}
L_{11}&0\\
0&L_{22}
\end{array}\right)
\end{equation*}
and, correspondingly,
\begin{equation*}
B^q=\left(\begin{array}{cc}
B_{11}^q&0\\
0&0
\end{array}\right)
\end{equation*}
where $L_{11}\in R^{s\times s} (0 \leq s < n)$ is the Laplacian
associated with those components connected with the virtual node,
$B_{11}^q\in R^{s\times s}$ is a nonzero diagonal matrix and $0$'s
denote some appropriate zero matrices.

Let $x = (y_1^T , y_2^T)^T$ with $y_1\in R^{sm}; y_2 \in
R^{(n-s)m}$. Then (\ref{eq2}) become:
\begin{equation*}
\begin{array}{ll}
\dot{y}_1&=-(H_1\otimes I_m)y_1+[B_1(I_k\otimes \textbf{1}_s)]\otimes I_m x_0,\\
\dot{y}_2&=-(L_{22}\otimes I_m)y_2,
\end{array}
\end{equation*}
where $B_{1}=[B_{11}^1,\cdots,B_{11}^k],
H_1=L_{11}+B_{1}(I_k\otimes \textbf{1}_s)$. From the proof for the
sufficient condition, it follows that $s$ agents corresponding to
$y_1$ will approach to the leaders for an arbitrary initial
conditions. However, the other agents corresponding to $y_2$ will
stay static or diverge to some distinct locations which can
arbitrarily exist and may not belong to the polytope
$co\{x_0^1,\cdots,x_0^k\}$.  This leads to a contradiction.
\hfill\rule{4pt}{8pt}

\begin{remark}
If $k=1$, then Theorem \ref{thm1} will be consistent with Theorem 4 in \cite{hong}.
\end{remark}

\section{SWITCHED INTERCONNECTION TOPOLOGY}

In this section, we consider the system (\ref{eq1}) associated
with switched interconnection topology. A result is given as
follows.

\begin{thm}
\label{thm2} For the system (\ref{eq1}), if $\bar{\mathcal{G}}_p
\;(p\in \mathcal{P})$ is connected, then (\ref{aim}) holds.
\end{thm}

Proof: Since the topology $\bar{\mathcal{G}}_p\;(p\in
\mathcal{P})$ is connected, from Lemma \ref{lem2}, $H_p\;(p\in
\mathcal{P})$ keeps positive definite, and moreover,
$[H_p^{-1}B_p(I_k\otimes \textbf{1}_n)]\otimes I_m$ is a row
stochastic matrix. Obviously, $[H_p^{-1}B_p(I_k\otimes
\textbf{1}_n)]\otimes I_m x_0\in \Xi$, from the differential
inclusion (\ref{eq1}), we have
\begin{equation}
\label{eq4}
\dot {x}\in \tilde{F}(x)=\{-(H_p \otimes I_m) (x-\xi)|\xi\in \Xi,p\in \cal P\}.
\end{equation}

Consider the Dini derivative of $d_\Xi(x(t))$,
\begin{equation}
\begin{array}{ll}
D^{+}d_\Xi(x(t))&=\underset{\xi\in \Xi}{inf}<x-\xi,\dot{x}>\\
                &\leq -\lambda_1 d_\Xi(x(t)),
\end{array}
\end{equation}
where $\lambda_1$ is the smallest (positive) eigenvalue of all
possible $H_p$ for $p\in \mathcal{P}$. Hence, $d_\Xi(x(t))$ is a
decreasing nonnegative function and $d_\Xi(x(t))=0$ if and only if
$x(t)\in \Xi$, then we have $d_\Xi(x(t))\to 0$ as $t\to
\infty$.\hfill\rule{4pt}{8pt}

\section{NUMERICAL EXAMPLES}

In this section, we give numerical simulations to illustrate the
coordination of multi-agent systems with multiple leaders. The
following two examples are considered (see Fig. 1 and Fig. 2).

\begin{center}
\includegraphics[width=2.5in]{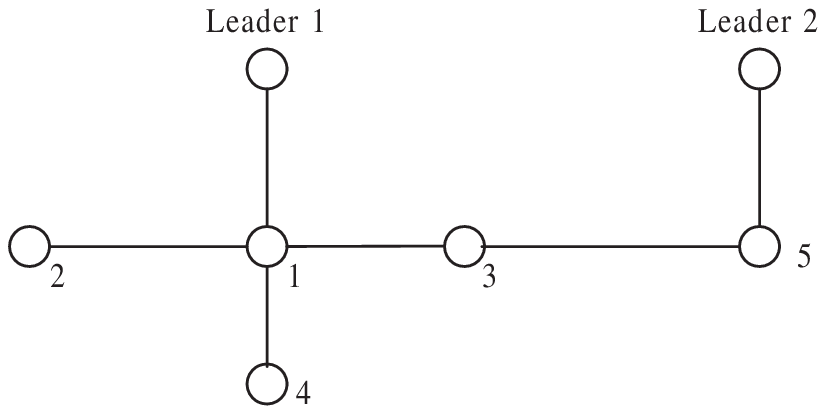}
\hspace{-0cm}\scriptsize
\begin{tabular}{c}
Fig.1.  $\bar{\mathcal{G}}_1$ and $\mathcal{G}_1$\\
\end{tabular}

\includegraphics[width=2.5in]{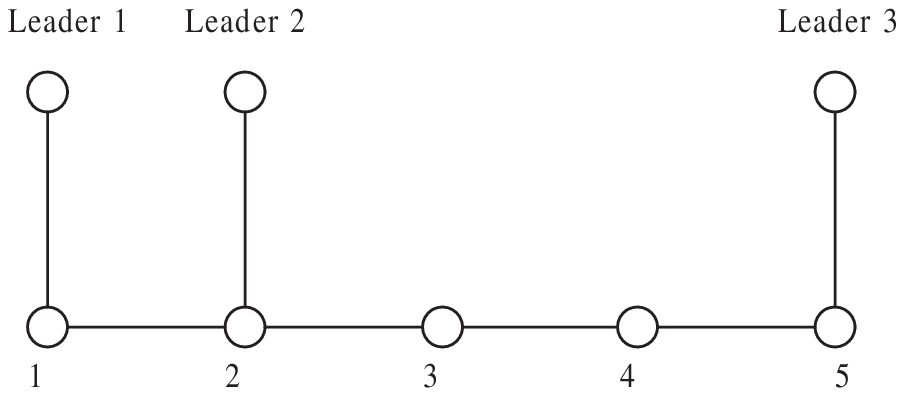}
\hspace{-0cm}\scriptsize
\begin{tabular}{c}
Fig.2.  $\bar{\mathcal{G}}_2$ and $\mathcal{G}_2$\\
\end{tabular}
\end{center}

In the first example, we take $n=5,k=2,m=1,t_0=0$ and the initial
positions of agents and leaders are given as follows:
$$
x_1(0)=5,\; x_2(0)=5.5, \; x_3(0)=6,\; x_4(0)=7,\;
$$
$$
x_5(0)=6.5, \; x_0^1=1,\; x_0^2=2.
$$
Then the simulation results show that the group of agents will
approach to the segment connecting the two leaders in Figs $3-6$.
In Fig. $4$, when the number of agents linked to leader $1$
increases, the agent group will move to leader 1 more closely.
Moreover, in Fig. 6, when agent $5$ is sensed by other agents,
then it also moves to leader 1. In Fig. 5, though agent 2 is not
connected with other agents but linked to leader 1, then it will
reach the locality of leader 1 finally.

\begin{center}
\includegraphics[width=2.5in]{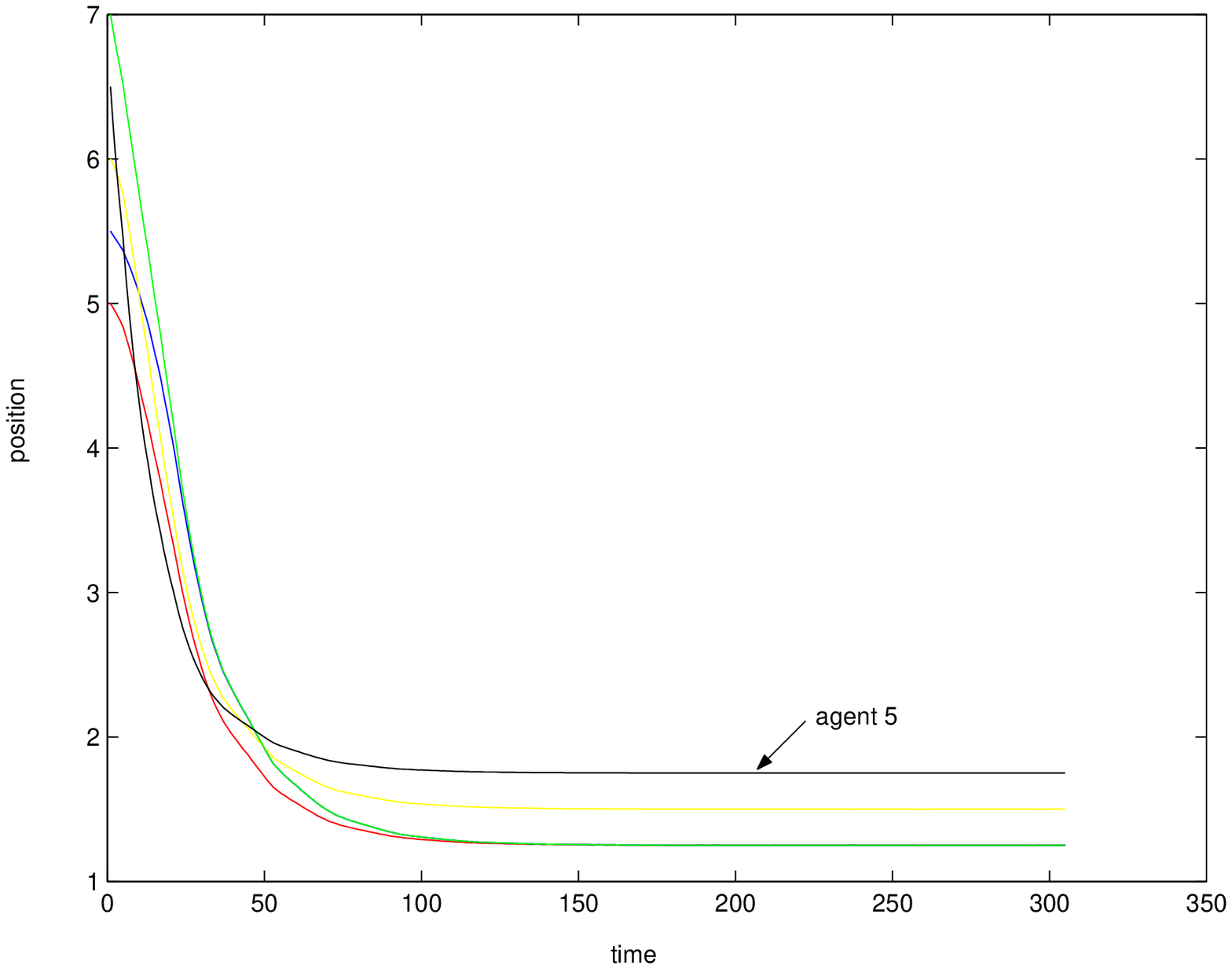}
\hspace{-0cm}\scriptsize
\begin{tabular}{c}
Fig.3. Coordination behavior withe the topology shown in Fig. 1\\
\end{tabular}

\includegraphics[width=2.5in]{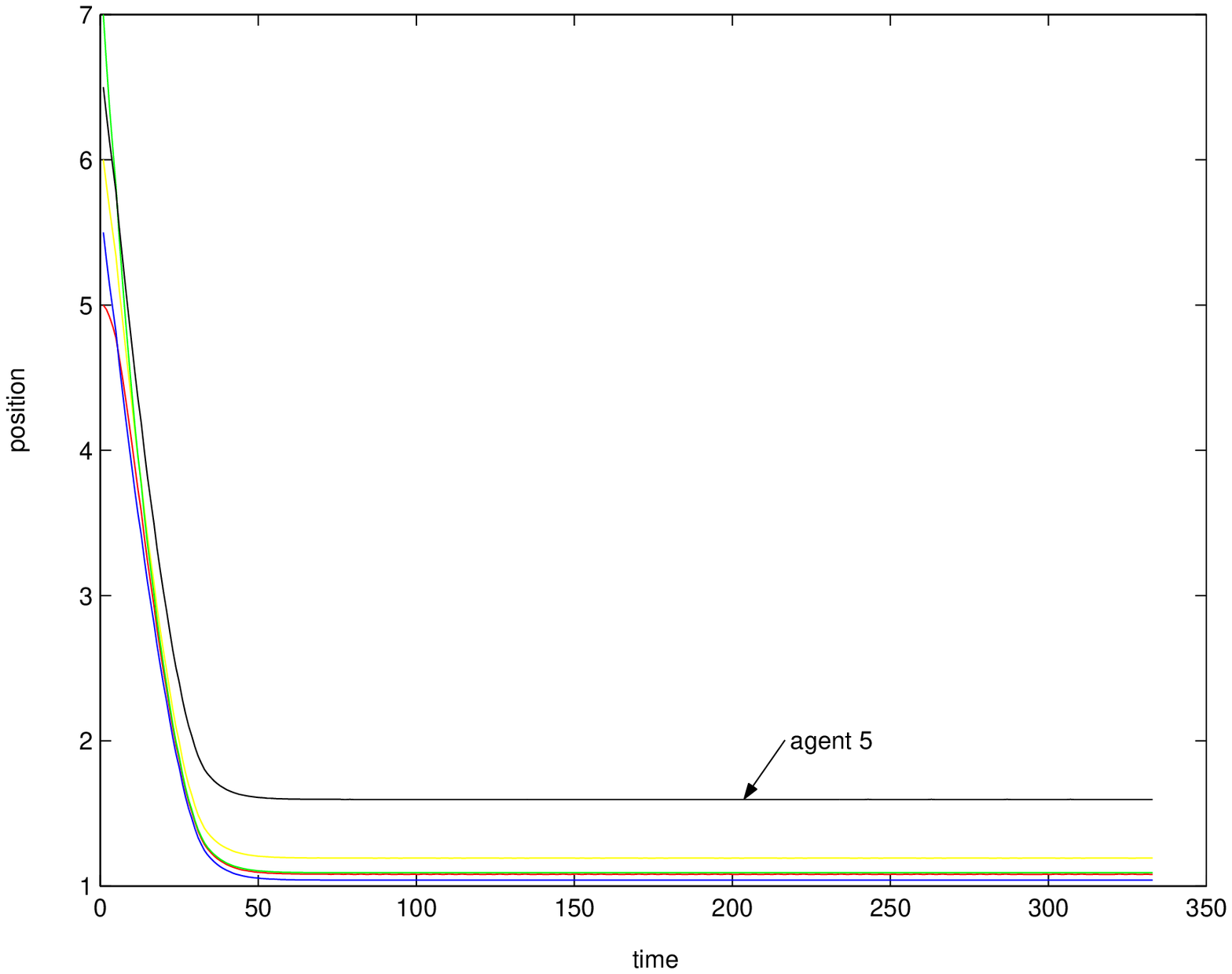}
\hspace{-0cm}\scriptsize
\begin{tabular}{c}
Fig.4. Coordination behavior based on revised Fig. 1, \\
where agents 2,3 and 4 connected with leader 1 \\
\end{tabular}
\end{center}

\begin{center}
\includegraphics[width=2.5in]{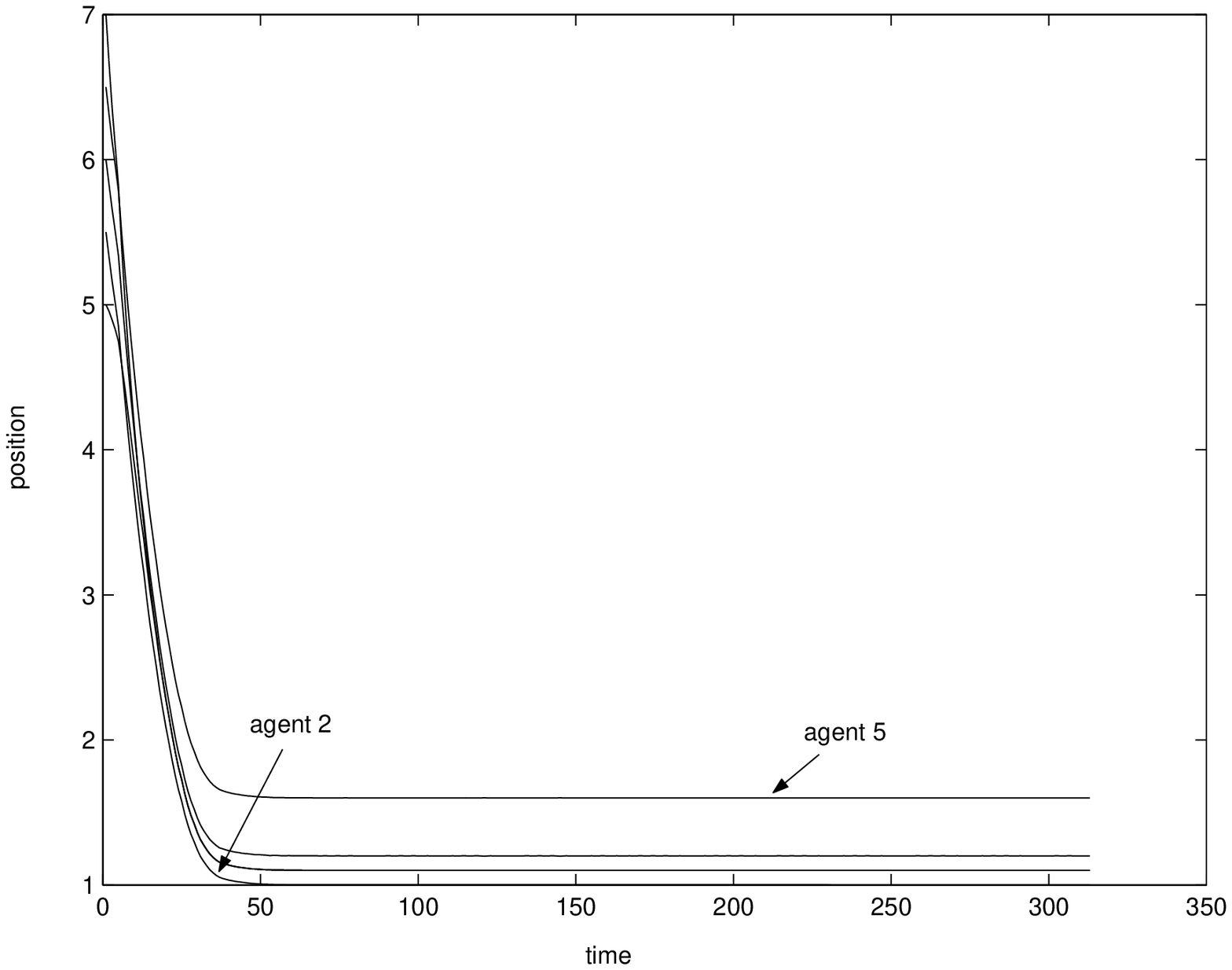}
\hspace{-0cm}\scriptsize
\begin{tabular}{l}
Fig.5. Coordination behavior with the topology\\
associated shown in Fig. 3 with agent 2 \\
disconnected with agents 3 and 4\\
\end{tabular}

\includegraphics[width=2.5in]{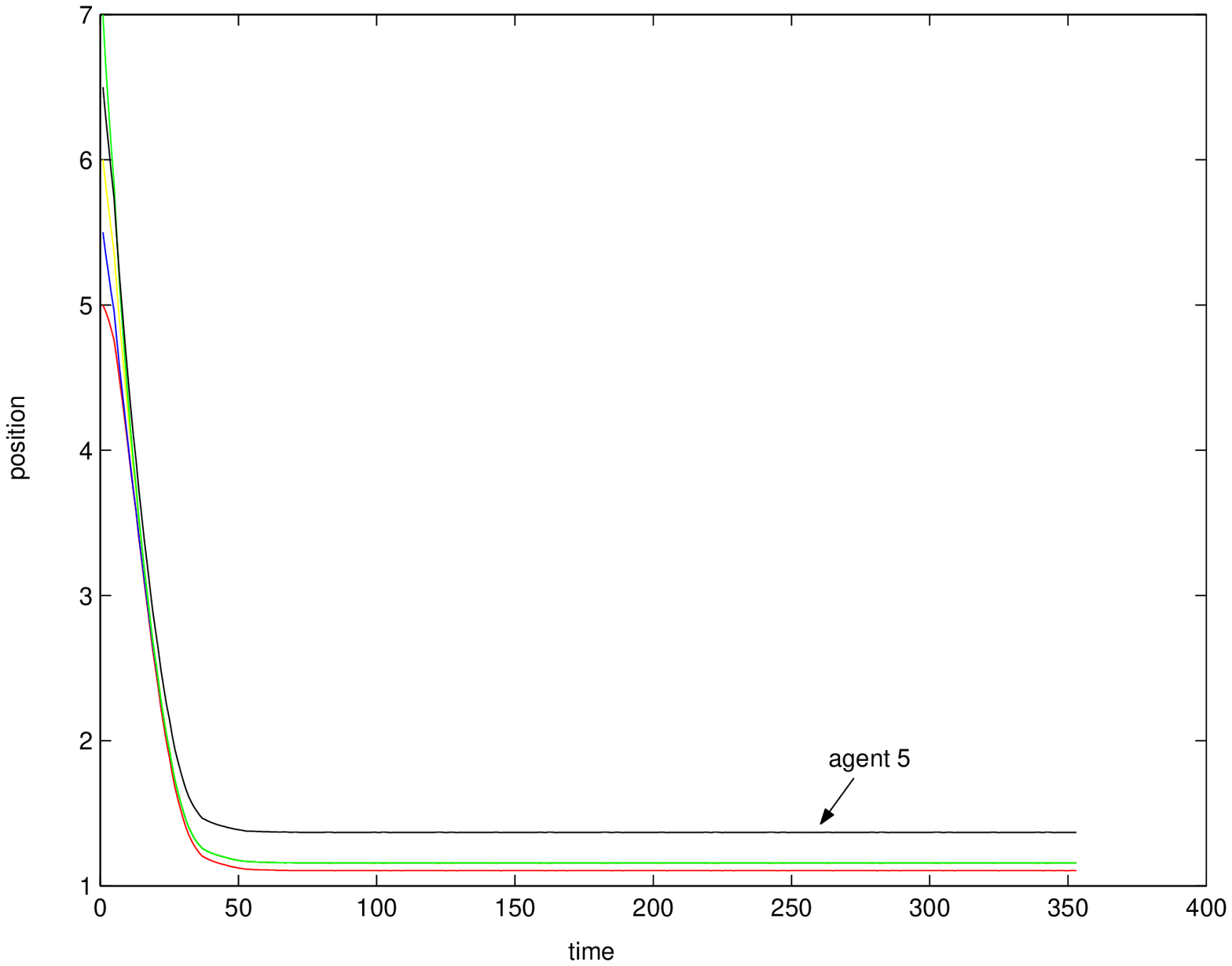}
\hspace{-0cm}\scriptsize
\begin{tabular}{l}
Fig.6. Coordination behavior with the topology\\
associated shown in Fig. 3 with agents 2 and 4\\
connected with agent 5 \\
\end{tabular}
\end{center}

In the second simulation example, we take $n=5,k=3,m=2,t_0=0$ and
the initial positions of agents and leaders are given as follows:
$$
x_1(0)=(0,0)^T,\; x_2(0)=(1,0)^T,\; x_3(0)=(2,0)^T,\;
$$
$$
x_4(0)=(3,0)^T,\; x_5(0)=(4,0)^T,\; x_0^1=(1,1)^T,\;
$$
$$
x_0^2=(2,2)^T,\; x_0^3=(1,2)^T.
$$
Here, the agents take a straight-line formation at the beginning.
\begin{center}
\includegraphics[width=2.5in]{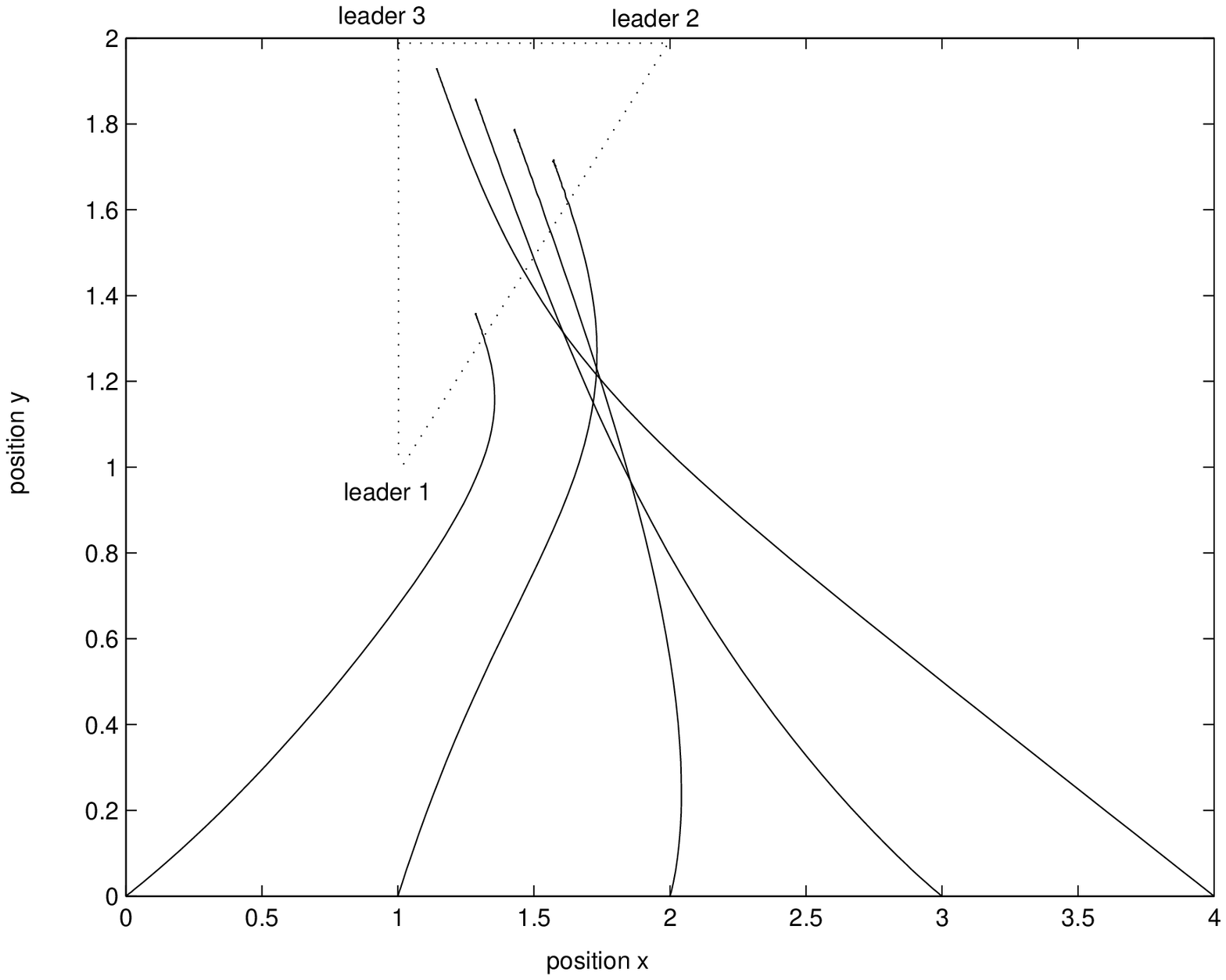}
\hspace{-0cm}\scriptsize
\begin{tabular}{l}
Fig.7. Coordination behavior related to Fig. 2
\end{tabular}

\end{center}

From Fig. 7, we observe that this group of agents will enter the
triangle formed by three leaders. Moreover, during the evolution,
agents 2, 3, 4, and 5 still remain straight-line formation.

\section{CONCLUSIONS}
\balance

This paper addressed a coordination problem of a multi-agent
system with multiple leaders. This group of agents were shown to
approach to the region ``contained" by the leaders if the
interconnection graph is connected.  Two interconnection cases,
fixed topology and switched topology, were discussed. Moreover,
numerical simulations were given to illustrate the theoretical
analysis.

\end{document}